\documentclass[12pt]{article}
\usepackage{epsfig,psfig}

\def\be{\begin{equation}}
\def\ee{\end{equation}}
\def\bc{\begin{center}}
\def\ec{\end{center}}
\def\bea{\begin{eqnarray}}
\def\eea{\end{eqnarray}}

\def\gappeq{\mathrel{\rlap {\raise.5ex\hbox{$>$}} {\lower.5ex\hbox{$\sim$}}}}
\def\lappeq{\mathrel{\rlap{\raise.5ex\hbox{$<$}} {\lower.5ex\hbox{$\sim$}}}}
\def\PR{{\it Phys.~Rev.~}}
\def\PRL{{\it Phys.~Rev.~Lett.~}}
\def\NP{{\it Nucl.~Phys.~}}

\def\PL{{\it Phys.~Lett.~}}

\catcode`@=11
\def\marginnote#1{}
\newcount\hour
\newcount\minute
\newtoks\amorpm
\hour=\time\divide\hour by60
\minute=\time{\multiply\hour by60 \global\advance\minute by-\hour}
\edef\standardtime{{\ifnum\hour<12 \global\amorpm={am}%
        \else\global\amorpm={pm}\advance\hour by-12 \fi
        \ifnum\hour=0 \hour=12 \fi
        \number\hour:\ifnum\minute<10 0\fi\number\minute\the\amorpm}}
\edef\militarytime{\number\hour:\ifnum\minute<10 0\fi\number\minute}
\def\draftlabel#1{{\@bsphack\if@filesw {\let\thepage\relax
   \xdef\@gtempa{\write\@auxout{\string
      \newlabel{#1}{{\@currentlabel}{\thepage}}}}}\@gtempa
   \if@nobreak \ifvmode\nobreak\fi\fi\fi\@esphack}
        \gdef\@eqnlabel{#1}}
\def\@eqnlabel{}
\def\@vacuum{}
\def\draftmarginnote#1{\marginpar{\raggedright\scriptsize\tt#1}}
\def\draft{\oddsidemargin 0.0truein
        \def\@oddfoot{\sl preliminary draft \hfil
        \rm\thepage\hfil\sl\today\quad\militarytime}
        \let\@evenfoot\@oddfoot \overfullrule 3pt
        \let\label=\draftlabel
        \let\marginnote=\draftmarginnote
   \def\@eqnnum{(\theequation)\rlap{\kern\marginparsep\tt\@eqnlabel}%
\global\let\@eqnlabel\@vacuum}  }
\catcode`@=12
\begin{document}
\begin{titlepage}
\vspace*{-1cm}
\phantom{hep-ph/0210323}
 
\hfill{DFPD-02/TH/23}
\hfill{CERN-TH/2002-250}
\hfill{SACLAY-T02/138}

\vskip 1.5cm
\begin{center} {\Large Models of Neutrino Masses:\\
\vskip .1cm Anarchy versus Hierarchy}
\end{center}
\vskip 1.  cm

\begin{center} {\large Guido Altarelli}~
\footnote{e-mail address: guido.altarelli@cern.ch}
\\
\vskip .1cm Theory Division, CERN,
\\  CH-1211 Gen\`eve 23, Switzerland
\\
\vskip .2cm {\large Ferruccio Feruglio}~\footnote{e-mail address: feruglio@pd.infn.it},
\\
\vskip .1cm Dipartimento di Fisica `G.~Galilei', Universit\`a di Padova and
\\  INFN, Sezione di Padova, Via Marzolo~8, I-35131 Padua, Italy
\\
\vskip .2cm {\large Isabella Masina}~\footnote{e-mail address: masina@spht.saclay.cea.fr},
\\
\vskip .1cm Service de Physique Th\'eorique, CEA-Saclay
\\  F-91191 Gif-sur-Yvette, France
\end{center}

\vskip 1.5cm
\begin{abstract}
\noindent We present a quantitative study of the ability of models with different levels of hierarchy to reproduce the solar
neutrino solutions, in particular the LA solution. As a flexible testing ground we
consider models based on SU(5)$\times$U(1)$_{\rm F}$. In this context, we have made statistical simulations of models with
different patterns from anarchy to various types of hierachy: normal hierarchical models with and without automatic
suppression of the 23 (sub)determinant and inverse hierarchy models.  We find that, not only for the LOW or VO solutions,
but even in the LA case, the hierarchical models have a significantly better success rate than those based on anarchy. The
normal hierachy and the inverse hierarchy models have comparable performances in models with see-saw dominance, while the
inverse hierarchy models are particularly good in the no see-saw versions. As a possible distinction between these
categories of models, the inverse hierarchy models favour a maximal solar mixing angle and their rate of success drops
dramatically as the mixing angle decreases, while normal hierarchy models are far more stable in this respect.

\end{abstract}
\end{titlepage}
\setcounter{footnote}{0}
\vskip2truecm

%
%

%
\section{Introduction} 
At present there are many possible models of $\nu$ masses and mixing \cite{af02}. This variety is mostly due to the
considerable experimental ambiguities that remain. In particular different solutions for solar neutrino oscillations
are still possible. Although the Large Angle (LA) solution emerges as the most likely from present data, the other
solutions LOW and Vacuum Oscillations (VO) are still not excluded (the small angle solution is very disfavoured by
now and we will disregard it in most of the following discussion). Indeed no solution is actually leading to an
imposingly good fit and, for example, the discrimination between LA and LOW is only based on a few hints which are
far from compelling \cite{strumia}. Hopefully in a few months, when the first results from the KamLAND experiment \cite{kamland}
will be known, one will have decisive evidence on this matter. Here we tentatively
disregard the possibility of a third neutrino oscillation frequency as indicated by the LSND
experiment \cite{lsnd} but not confirmed by KARMEN \cite{karmen} and to be finally checked by the MiniBOONE experiment \cite{miniboone} now about to start.

For model building there is an important quantitative difference between the LA solution on the one side and the LOW
or VO solutions on the other side. While for all these solutions the relevant mixing angle
$\theta_{12}$ is large, the value of the squared mass difference $\Delta m^2_{12}=m_2^2-m_1^2$ (with, by definition,
$m_2^2\ge m_1^2$) is very different for LA, LOW and VO: $\sim 10^{-4} ~eV^2$, 
$\sim 10^{-7} ~eV^2$ and $\sim 10^{-10} ~eV^2$, respectively. Thus the gap with respect to the atmospheric neutrino
oscillation frequency $\Delta m^2_{23}=m_3^2-m_2^2$, which is given by
$|\Delta m^2_{23}|\sim 3\cdot 10^{-3} ~eV^2$, is moderate for LA and very pronounced for the other two solutions.

For
the LOW and VO solutions the large frequency difference with respect to that of atmospheric neutrinos points to a
hierarchical spectrum for the three light neutrinos. Possible hierarchical patterns are the normal hierarchy case
$m_3\gg m_2\gappeq m_1$ or the inverted hierarchy alternative $m_2\gappeq m_1\gg m_3$ (in this case $\Delta
m^2_{23}$ is negative in our definition). Then a main problem is to explain the presence of large mixing angles
between largely splitted mass states (in particular the atmospheric neutrino oscillation mixing angle $\theta_{23}$
is experimentally close to maximal). In hierarchical models the consistency of these usually opposed constraints is
obtained  by mechanisms that guarantee a vanishing or a strongly suppressed 23 sub-determinant. In the see-saw
mechanism for neutrino masses this suppression can be naturally obtained, for example, in the so-called lopsided
models \cite{lopsided} and/or by the dominance \cite{dominance} of one eigenvalue in
$M^{-1}$, $M$ being the right-handed (RH) Majorana matrix. Models of this type
have been studied and provide, as also quantitatively confirmed by our present analysis, an essentially unique
framework for a successful description of both atmospheric and solar neutrino oscillations when the LOW or the VO
solutions are adopted.

In the case of the LA solution the ratio of the solar and atmospheric $\Delta m^2$ ranges is typically given by
\be r=\Delta m^2_{12}/|\Delta m^2_{23}|\sim 1/20-1/100.  \label{r}
\ee For LA one can reproduce the data either in a nearly degenerate or in a hierarchical model. In a degenerate
model, due to laboratory and cosmological bounds, the  common value of $m_i^2\sim m^2$ cannot exceed a few $~eV^2$.
But the actual value is probably well below this level because of the constraint imposed by neutrinoless double beta
decay ($0\nu \beta \beta$) \cite{doublebeta} that would otherwise require a strong cancellation, only possible for nearly maximal
solar oscillation mixing \cite{cancellation}. This fact, together with the general difficulty, in the absence of a specific mechanism,
of obtaining too small values of $\Delta m^2_{12}/m^2$, suggests that a moderate degeneracy is more likely. 
Typically we could have all
$m_i^2 \sim ({\rm few}~10^{-3}-10^{-2})~eV^2$ with one comparatively not-so-small splitting $\Delta m^2_{12}\sim
10^{-4}~eV^2$. Or, as a different example, we can have a (normal) hierarchical model with $m_3^2\sim 3\cdot 10^{-3}
~eV^2$,
$m_2^2\lappeq 10^{-4} ~eV^2$ and $m_1^2\sim 0$ or an (inverse) hierarchical model with $m_{1,2}^2\sim 3\cdot 10^{-3}
~eV^2$, and $m_3^2\sim 0$. Actually a sufficient hierarchy (a factor of ~5 in mass) can arise from no significant
underlying structure at all. In particular, the see-saw mechanism, being quadratic in the Dirac neutrino masses,
tends to enhance small fluctuations in the Dirac  eigenvalue ratios. This is the point of view of anarchical models \cite{anarchy1},
where no structure is assumed to exist in the neutrino mass matrix and the smallness of $r$ is interpreted as a
fluctuation. But one additional feature of the data plays an important role in this context and presents a clear
difficulty for anarchical models. This is the experimental result that the third mixing angle
$\theta_{13}$ is small, $\vert U_{e3}\vert=\vert\sin{\theta_{13}}\vert\lappeq 0.2$ \cite{chooz}. So, for neutrinos two mixing
angles are large and one is small. Instead in anarchical models all angles should apriori be comparable and not
particularly small. Therefore this is a difficulty
for anarchy and, for the survival of these models, it is necessary that $\theta_{13}$ is found very
close to the present upper bound. Instead in hierarchical models the smallness of $\theta_{13}$ can be
obtained as a reflection of the underlying structure in that some small parameter is present from the beginning in
these models.

In this note we make a quantitative study of the ability of different models to reproduce the solar neutrino
solutions. As a flexible testing ground we consider models based on SU(5)$\times$U(1)$_{\rm F}$.  The SU(5)
generators  act ``vertically'' inside one generation, while the U(1)$_{\rm F}$ charges are different
``horizontally'' from one generation to the other. If, for a given interaction vertex, the U(1)$_{\rm F}$ charges do
not add to zero, the vertex is forbidden in the symmetric limit. But the symmetry is spontaneously broken by the VEV
$v_f$ of a number of ``flavon'' fields with non vanishing charge. Then a forbidden coupling is rescued but is
suppressed by powers of the small parameters $\lambda_f=v_f/\Lambda$ with the exponent larger for larger charge 
mismatch \cite{fn}. We expect
$v_f\gappeq M_{GUT}$ and, for the cut-off $\Lambda$ of the theory,
$\Lambda \lappeq M_{Pl}$.  In these models \cite{su5timesu1,othercharges} the known generations of quarks and leptons are contained in triplets
$\Psi^{10}_i$ and
$\Psi^{\bar 5}_i$, $(i=1,2,3)$ corresponding to the 3 generations, transforming as $10$ and ${\bar 5}$ of SU(5),
respectively. Three more SU(5) singlets
$\Psi^1_i$ describe the RH neutrinos. In SUSY models we have two Higgs multiplets, which transform as 5 and $\bar 5$
in the minimal model. All mass matrix elements are of the form of a power of a suppression factor times a number of
order unity, so that only their order of suppression is defined. We restrict for simplicity to integral charges:
this is practically a forced choice for the LA case where the hierarchy parameter must be relatively large (so that
$\sqrt{\lambda}\sim 0(1)$), while for the LOW and VO cases, where the hierarchy parameter is small, it is only
motivated by the fact that enough flexibility is obtained for the present indicative purposes. There are
many variants of these models \cite{af02}: fermion charges can all be non negative with only negatively charged flavons, or
there can be fermion charges of different signs with either flavons of both charges or only flavons of one charge.
The Higgs charges can be equal, in particular both vanishing or can be different. We will make use of this
flexibility in order to study the relative merits of anarchy versus various degrees and different patterns of
hierarchy. 

In this context we have studied in detail different classes of models: normal hierarchical models with and
without automatic suppression of the 23 (sub)determinant, inverse hierarchy models and anarchical models. The normal
hierarchical models without automatic suppression of the 23 determinant are clearly intermediate: in
a sense in those cases anarchy is limited to the 23 sector. We denote them as partially hierarchical or semi-anarchical in
the following.  We also compare, when applicable, models with light neutrino masses dominated by the see-saw mechanism or
by non renormalizable dim-5 operators. We
construct our models by assigning suitable sets of charges for
$\Psi^{10}_i$,
$\Psi^{\bar 5}_i$ and $\Psi^1_i$. In all input mass matrices the coefficients multiplying the power of the hierarchy
parameter are generated at random as real and complex numbers in a given range of values
\cite{anarchy1,random}. We compare the case of real or
complex parameters and we also discuss the delicate questions of the
probability distribution for the coefficients and the stability of the results. We assign a merit factor to each model
given by the percentage of success over a large sample of trials. For each model the value of the hierarchy parameter is
adjusted by a coarse fitting procedure to get the best rate of success. 

Our results can be summarized as follows. As expected, for the LOW and VO cases only
hierarchical model provide a viable approach: in comparison the rate of success of anarchical and seminarchical models is
negligible. But also for the LA solution we still find that hierarchical models are sizeably better in general. The most
efficient ones are inverse hierarchy models with no see-saw dominance, which are more than 10 times better with respect
to anarchy with see-saw (anarchy prefers the see-saw case by about a factor of 2). Among the see-saw dominance versions
the most performant models remain the hierarchical ones (by a factor
of about 4 with respect to anarchy with see-saw) with not much difference between inverse or normal hierarchy.
Semi-Anarchical models are down by a factor of about 2 with respect to hierarchical models among the see-saw versions (but
this value is less stable with respect to changes of the extraction procedure and, for example, it tends to be washed out
going from complex to real coefficients). In all models the $\tan^2{\theta_{23}}$ distribution is in agreement with
large mixing but it is not sharply peaked around 1 as for maximal mixing. Near maximal mixing is instead a
prediction for solar neutrinos in inverse hierarchical models, so that their advantage with respect to other models
would be rapidly destroyed if the data will eventually move in the direction away from maximality.

%
%

\vspace{0.5cm}
\section{Framework}

We consider a class of models with an abelian flavour symmetry compatible with SU(5) grand unification. 
Here we will not address the well-known problems of grand unified theories, such as the doublet-triplet
splitting, the proton lifetime, the gauge coupling unification beyond leading order and the wrong mass relations for
charged fermions of the first two  generations. We adopt the SU(5)$\times$U(1)$_{\rm F}$ framework simply as a convenient
testing ground for different neutrino mass scenarios. In all the models that we study the large atmospheric mixing angle
is described by assigning equal flavour charge to muon and tau neutrinos and their weak SU(2) partners (all belonging
to the 
${\bar 5}\equiv(l,d^c)$ representation of SU(5)).  Instead, the solar neutrino oscillations can be obtained with
different, inequivalent charge assignments and both the LOW\footnote{The LOW and VO solutions can be fitted almost
equally well by suitable models. Thus here we will focus mainly on the LOW case.} and the LA solution can be
reproduced. 

A first class of models is characterized by all matter fields having  flavour charges of one sign, for example
all non negative. An important property
of models in this class is that the light neutrino mass matrix $L^Tm_{\nu}L$ is independent from
the charges of both the 10
$\equiv(q,u^c,e^c)$ and 1
$\equiv \nu^c$ representations , even in the see-saw case when $m_{\nu}=m_D^TM_R^{-1}m_D$. For $m_{\nu}$ entries the 
powers of the symmetry breaking parameter
$\lambda$ are dictated by the charges F of the ${\bar 5}$.  Since in this case what really matters are charge
differences, rather than absolute values, the equal charges for the second  and third generations can be put to zero,
without loosing generality:
\be {\rm F}({\bar 5})=(b,0,0)~~~~~~~~~~~~~b\ge 0~~~.
\label{chargfb}
\ee If $b$ also vanishes, then the light neutrino mass matrix will be structure-less and we will call anarchical (A)
this sub-class of models. In a large sample of anarchical models, generated with random coefficients, the resulting
neutrino mass spectrum can exhibit either normal or inverse hierarchy. Anarchical models clearly prefer the LA
solution with a moderate separation between atmospheric and solar frequencies. They tend to favour large, not
necessarily maximal, mixing angles,  including $U_{e3}$, which represents a problem. Therefore,
in anarchical models, $U_{e3}$ is expected 
to be close to the present experimental bound.

If $b$ is positive, then the light neutrino mass matrix will be structure-less only in the (2,3) sub-sector and we
will call semi-anarchical (SA) the corresponding models. In this case, the neutrino mass spectrum has normal hierarchy. However,
unless the (2,3) sub-determinant is accidentally suppressed, atmospheric and solar oscillation frequencies are 
expected to be of the same order and, in addition, the preferred solar mixing angle is small. Nevertheless, such a
suppression can occur in a fraction of semi-anarchical models generated with random, order one coefficients. The
real advantage over the fully anarchical scheme is represented by the suppression in $U_{e3}$. 

In a second class of models matter fields have both positive and negative flavour charges. In these models, the
light neutrino mass matrix will in general depend also on the charges of 10 and 1. A first sub-case arises when only
the RH neutrino fields have charges of both signs.  It has been shown that it is possible to exploit this
feature to reproduce a neutrino mass spectrum with normal hierarchy and a natural gap between atmospheric and solar
frequencies. Via the see-saw mechanism the (2,3) sub-determinant  vanishes in the flavour symmetric limit. At the
same time a large solar mixing angle can be obtained. Clearly this is particularly relevant for the LOW (or VO)
solution. It is less clear to which extent the condition of vanishing determinant  is needed for the LA solution and
one of the purposes of the present paper is precisely to compare these models, which we will call hierarchical (H) with the
anarchical and semi-anarchical models, that do not reproduce  such a condition.

Finally, we can have fields with charges of both signs in both the 1 and the
${\bar 5}$. In this context it is possible to reproduce an inverse hierarchical spectrum, with a large (actually,
almost maximal) solar mixing angle and a large gap between atmospheric and solar frequencies
\cite{invhier}. Also this sub-class of
models, which we call inversely hierarchical (IH), are appropriate to describe  both LOW and   LA solutions. 
\vspace{0.1cm}
\begin{table}[!h]
\caption{Models and their flavour charges.
\label{tab1}}
\vspace{0.4cm}
\begin{center}
\begin{tabular}{|c|c|c|c|c|}
\hline & & & & \\ {\tt Model}& ${{\Psi_{10}}}$ & ${\Psi_{\bar 5}}$ & ${{\Psi_1}}$ & ${(H_u,H_d)}$ \\ & & & & \\
\hline
\hline & & & & \\ {\tt Anarchical (A)}& (3,2,0)& (0,0,0) & (0,0,0) & (0,0)\\ & & & & \\
\hline & & & & \\ {\tt Semi-Anarchical (SA)}& (2,1,0) & (1,0,0) & (2,1,0) & (0,0) \\ & & & & \\
\hline
\hline & & & & \\ {\tt Hierarchical (H)}& (3,2,0)& (2,0,0) & (1,-1,0) & (0,0)\\ & & & & \\
\hline & & & & \\ {\tt Inversely Hierarchical (LA)}& (3,2,0)& (1,-1,-1)& (-1,+1,0)& (0,+1)\\ & & & & \\
\hline & & & & \\ {\tt Inversely Hierarchical (LOW)}& (2,1,0)& (2,-2,-2)& (-2,+2,0)& (0,+2)\\ & & & & \\
\hline
\end{tabular}
\end{center}
\end{table}

The hierarchical and the inversely hierarchical models may come into  several varieties depending on the number and
the charge of  the flavour symmetry breaking (FSB) parameters. Here we will consider both the case (I) of a single,
negatively charged flavon, with symmetry breaking parameter $\lambda$ or that of two (II)  oppositely charged flavons with symmetry breaking
parameters $\lambda$ and $\lambda'$. In case I, it is impossible to compensate negative F charges in the Yukawa
couplings and the corresponding entries in  the neutrino mass matrices vanish. Eventually these zeroes are filled by
small contributions, arising, for instance, from the diagonalization of the charged lepton sector or from the
transformations needed to make the kinetic terms canonical. In our analysis we will always include effects coming
from the charged lepton sector, whereas we will neglect those coming from non-canonical kinetic terms. 

Another important ingredient in our analysis is represented by the see-saw mechanism \cite{seesaw}. Hierarchical models and
semi-anarchical  models have similar charges in the $(10,{\bar 5})$ sectors and, in the absence of the see-saw
mechanism, they would give rise to similar results. Even when the results are expected to be independent from the
charges of the RH neutrinos, as it is the case for the anarchical and semi-anarchical models, the see-saw
mechanism can induce some sizeable effect in a statistical  analysis. For this reason, for each type of model, but
the hierarchical  ones (the mechanism for the 23 sub-determinant suppression is in fact based on the see-saw
mechanism), we will separately study the case where RH neutrinos are present and the case where
they are absent. When RH neutrinos are present, there are two independent  contributions to the light
neutrino mass matrix. One of them comes  via the see-saw mechanism from the exchange of the heavy RH  modes. The
other one is provided by L-violating dimension five operators arising from physics beyond the cut-off. These
contributions have the same transformation properties under the flavour group and, in general, add coherently. In
our analysis we will analyze the case where the see-saw contribution is the dominant one (SS). The absence of RH
neutrinos describes the opposite case, when the mass matrix is saturated by the non-renormalizable contribution
(NOSS). 

For each type of model we have selected what we consider to be a typical representative 
\footnote{We made no real optimization effort to pick up the `most' representative, but rather a model with a high
success rate in its class.} and we have collected in table 1 the corresponding charges.  In the next section we will
compare the performances of the following models:
${\rm A_{SS}}$, ${\rm A_{NOSS}}$, ${\rm SA_{SS}}$, ${\rm SA_{NOSS}}$,
${\rm H_{(SS,I)}}$, ${\rm H_{(SS,II)}}$, ${\rm IH_{(SS,I)}}$, 
${\rm IH_{(SS,II)}}$, ${\rm IH_{(NOSS,I)}}$ and ${\rm IH_{(NOSS,II)}}$.
\vspace{0.1cm}
\begin{table}[!h]
\caption{Order of magnitude predictions for oscillation parameters, from neutrino mass
matrices in eq. (\ref{mnuh}) and (\ref{mnuih}) \cite{af02}; $d_{23}$ denotes the sub-determinant in the 
23 sector and we show the effect of its accidental suppression for the semi-anarchical model. In the estimates 
we have chosen $\lambda=\lambda'$. Inverse hierarchy predicts an almost maximal $\theta_{12}$.
\label{tab2}}
\vspace{0.4cm}
\begin{center}
\begin{tabular}{|c|c|c|c|c|c|c|}
\hline & & & & & &\\ 
{\tt Model}& {\tt parameters}& $d_{23}$ & $\Delta m^2_{12}/\vert\Delta m^2_{23}\vert$& $U_{e3}$& $\tan^2\theta_{12}$
&$\tan^2\theta_{23}$\\ 
& & & & & &\\
\hline
& & & & & &\\
{\tt A} &$\epsilon=1$ & O(1) & O(1) & O(1) & O(1) & O(1) \\
& & & & & &\\
\hline
& & & & & &\\
{\tt SA} &$\epsilon=\lambda$ & O(1) & O($d_{23}^2$) & O($\lambda$) & O($\lambda^2/d_{23}^2$)  & O(1) \\
& & & & & &\\
\hline
& & & & & &\\
{\tt H$_{\tt II}$} &$\epsilon=\lambda^2$ & O($\lambda^2$) & O($\lambda^4$) & O($\lambda^2$) & 
O(1)  & O(1) \\
& & & & & &\\
\hline
& & & & & &\\
{\tt H$_{\tt I}$} &$\epsilon=\lambda^2$ & 0 & O($\lambda^6$) & O($\lambda^2$) & 
O(1)  & O(1) \\
& & & & & &\\
\hline
& & & & & &\\
{\tt IH (LA)} &$\epsilon=\eta=\lambda$ & O($\lambda^4$) & O($\lambda^2$) & O($\lambda^2$) & 
1+O($\lambda^2$)  & O(1) \\
& & & & & &\\
\hline
& & & & & &\\
{\tt IH (LOW)} &$\epsilon=\eta=\lambda^2$ & O($\lambda^8$) & O($\lambda^4$) & O($\lambda^4$) & 
1+O($\lambda^4$)  & O(1) \\
& & & & & &\\
\hline
\end{tabular}
\end{center}
\end{table}

Anarchical, semi-anarchical and hierarchical models give rise to a mass
matrix for light neutrinos of the type
\be
m_\nu=
\left(
\begin{array}{ccc}
\epsilon^2 & \epsilon &\epsilon\\
\epsilon & 1 & 1\\
\epsilon & 1 & 1
\end{array}
\right)~~~~~~~~~~({\tt A,SA,H})~~~,
\label{mnuh}
\ee
where all the entries are specified up to order one coefficients
and the overall mass scale has been conventionally set to one.
For anarchical models, $\epsilon=1$. Then all the entries are uncorrelated
numbers of order one and no particular pattern becomes manifest.
For the semi-anarchical model of table 1, $\epsilon=\lambda$. There
is a clear distinction between the first row and column and the 
23 block of the mass matrix, which is structureless as in the anarchical 
models. In particular, barring accidental cancellations, the sub-determinant
in the 23 sector is of order one. Finally, the hierarchical model
defined by the choice of charges in table 1, has $\epsilon=\lambda^2$.
At variance with the anarchical or semi-anarchical models, 
the determinant of the 23 sector is suppressed by the see-saw mechanism
and is of order $\lambda\lambda'$.

The inversely hierarchical models are characterized by a neutrino mass matrix
of the kind
\be
m_\nu=
\left(
\begin{array}{ccc}
\epsilon^2 & 1 & 1\\
1 & \eta^2 & \eta^2\\
1 & \eta^2 & \eta^2
\end{array}
\right)~~~~~~~~~~({\tt IH})~~~,
\label{mnuih}
\ee
where $\epsilon=\lambda$ $(\lambda^2)$ and $\eta=\lambda'$ $(\lambda'^2)$
for the LA (LOW) solution. The ratio between the solar and atmospheric
oscillation frequencies is not directly related to the sub-determinant 
of the block 23, in this case. 
The above mass matrices also receive an additional contribution from the
diagonalization of the charged lepton sector, which, however, does
not spoil the displayed structure. For completeness, we collect in table
2 the gross features of the models under consideration.
Notice that the hierarchical models predict a ratio $\Delta m^2_{12}/\vert\Delta m^2_{23}\vert$
of order $\lambda^4$ or $\lambda^6$. In these cases it is possible to fit both the LOW
and the LA solutions with an expansion parameter $\lambda\approx\lambda'$ that,
within a factor of two, matches the Cabibbo angle. On the contrary,
the inversely hierarchical model that reproduces the LA solution
favours $\Delta m^2_{12}/\vert\Delta m^2_{23}\vert\approx O(\lambda^2)$.
If we adopt this model to fit also the LOW solution, 
the corresponding values of $\lambda\approx\lambda'$ do not provide   
a decent description of the remaining fermion masses. For this reason, 
when analyzing the LOW solution, we have considered a separate set of charges 
for the inverted hierarchy.

%
%

\vspace{0.5cm}
\section{Method and results}
Abelian flavour symmetries predict each entry of fermion mass matrices up to an unknown dimensionless coefficient. These
coefficients, that are free-parameters of the theory, are expected to have absolute values of order one. Aside from
generalized kinetic terms, which we do not consider, the relevant mass matrices are specified by $N=24$ order one
parameters: 9 from the charged lepton sector, 9 from Dirac neutrino mass matrix entries and  6 from Majorana RH neutrino
mass matrices. When RH neutrinos are absent (that is, for the NOSS cases) the LH neutrino mass matrix contains 6 parameters
and the relevant set reduces to $N=15$ parameters. We have analyzed the case of real or complex coefficients
${\cal P}=({\cal P}_1,... {\cal P}_N)$, with absolute values generated as random numbers in an interval ${\cal I}$ and
random phases taken  in $[0,2\pi]$. To study the dependence of our results on ${\cal I}$, we have considered several
possibilities: $[0.5,2]$ (default),
$[0.8,1.2]$, $[0.95,1.05]$ and $[0,1]$. In the case of real coefficients, which is studied for comparison, we allow both
signs for the coefficients. For each model, only a portion
$V_S$ of the volume 
$V=(2 \pi{\cal I})^N$ of the parameter space gives rise to predictions in  agreement with the experimental data within the
existing uncertainties. We may interpret $V_S/V$ as the success rate of the model in describing neutrino data.  Clearly this
portion shrinks to zero for infinitely good measurements. Therefore we are not interested in its absolute size, but rather
in the relative sizes of $V_S/V$ in different models.

We evaluate the success rate of each model by considering, through a random generation, a large number of `points'
${\cal P}$ and by checking whether the corresponding predictions do or do not fall in the experimentally allowed regions \cite{random}. To
this purpose we perform a test based on four observable quantities:  $O_1=r
\equiv \Delta m^2_{12}/\vert\Delta m^2_{23}\vert$,
$O_2=\tan^2\theta_{12}$, $O_3=\vert U_{e3}\vert\equiv\vert\sin\theta_{13}\vert$ and 
$O_4=\tan^2\theta_{23}$. We take \cite{fits}:
\be
\begin{array}{l} 4\cdot 10^{-6} < r < 4\cdot 10^{-5}\\
\vert U_{e3} \vert < 0.2\\ 0.52 <\tan^2\theta_{12} < 1.17\\ 0.33 <\tan^2\theta_{23}<3.3
\end{array} ~~~~~~~~~{\rm (LOW)}
\label{low}
\ee
\be
\begin{array}{l} 0.01 < r < 0.2\\
\vert U_{e3} \vert < 0.2\\ 0.24 <\tan^2\theta_{12}<0.89\\ 0.33 <\tan^2\theta_{23}<3.3
\end{array} ~~~~~~~~~{\rm (LA)}
\label{la}
\ee The boundaries of these windows are close to the $3\sigma$ limits on the corresponding observable quantity. The test is
successfully passed if $O_i({\cal P})$ are in the above windows, for all $i=1,...,4$.  We will study the dependence on the
choice made in eqs. (\ref{low},\ref{la}), by analyzing the distributions of the points generated for each observable. We
then estimate $V_S/V$ from the ratio between the number of successful trials over the  total number of attempts. Notice that
we do not extend our test to the remaining fermion masses and mixing angles, though the  models considered can also
reproduce, at the level of order of magnitudes, charged lepton masses, quark masses and mixing angles. With the interval
${\cal I}$ fixed at its reference value, $[0.5,2]$, the different models are compared at their best performance, after
optimizing for each model the value of the symmetry breaking parameters
$\lambda$ and $\lambda'$.

\vspace{0.2cm}
\subsection{LOW solution}

\begin{figure}[!ht]
\centerline{\psfig{file=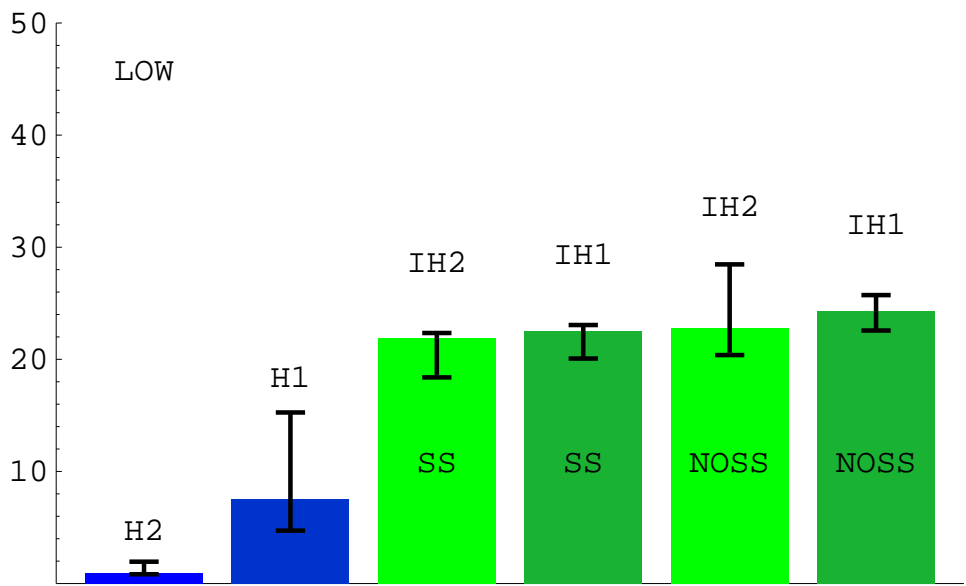,width=0.85\textwidth}}
\caption{Relative success rates for the LOW solution. The sum of the
rates has been normalized to 100. The results correspond to the 
default choice ${\cal I}=[0.5,2]$, and to the following values 
of $\lambda=\lambda'$: $0.1$, $0.15$, $0.03$, $0.04$, $0.05$, $0.06$
for the models ${\rm H_{(SS,II)}}$, ${\rm H_{(SS,I)}}$, ${\rm IH_{(SS,II)}}$, 
${\rm IH_{(SS,I)}}$, ${\rm IH_{(NOSS,II)}}$ and ${\rm IH_{(NOSS,I)}}$,
respectively. The error bars represent
the linear sum of the systematic error due to the
choice of ${\cal I}$ and the statistical error (see text).
The results for the A and SA models are below 0.01,
independently from ${\cal I}$ and from $(\lambda,\lambda')$,
and are not displayed.}
\label{flow}
\vskip 1cm
\end{figure}

In fig. \ref{flow} we compare the success rates for the LOW solution. In this figure the anarchical or semi-anarchical models do not appear
simply because their rates of success are negligible on the scale of the figure.

\vskip .5cm

It is clear that, if the future experimental results will indicate the LOW solution as the preferred one, then
anarchical or semi-anarchical schemes will be completely inadequate to describe the data.  It would be natural in that case
to adopt a model where the large gap between the atmospheric and the solar oscillation  frequencies is built in as a result
of a symmetry. What is striking about the results displayed in fig. \ref{flow} is the ability of the IH models, both
in the SS and in the NOSS versions, to reproduce the data. An appropriate choice for
$\lambda$ ($\lambda'$) is completely successful in reproducing
$r$. $U_{e3}$ is numerically close to $r$ and easily respects  the present bound. Moreover, $\tan^2 \theta_{12}$ is very
close to 1, due to the pseudo-Dirac structure of the 12 sector. Only the $\tan^2 \theta_{23}$ distribution shows some
flatness  and contributes to deplete the success rate. It is worth stressing that $\tan^2 \theta_{12}$ is so strongly
peaked around 1, that any significant deviation from $\tan^2 \theta_{12}=1$ in the data  would provide a severe difficulty
for the IH schemes. The H model has a smaller success rate (still much larger than that of the A and SA models), but has
smoother distributions for the four observables and is less sensitive than IH to  variations of the experimental data. The
error bars in fig. \ref{flow} are dominated by the systematic effects, which have been estimated by varying the interval
${\cal I}$. We considered four possibilities: ${\cal I}=[0.5,2]$  (default), ${\cal I}=[0.8,1.2]$, ${\cal I}=[0.95,1.05]$,
and 
${\cal I}=[0,1]$. To the highest (lowest) rate of each model we then add (subtract) linearly the statistical error.  The
latter is usually smaller than the systematic error. Further details are reported in appendix A.

\vspace{0.5cm}
\subsection{LA solution}

\begin{figure}[!h]
\centerline{\psfig{file=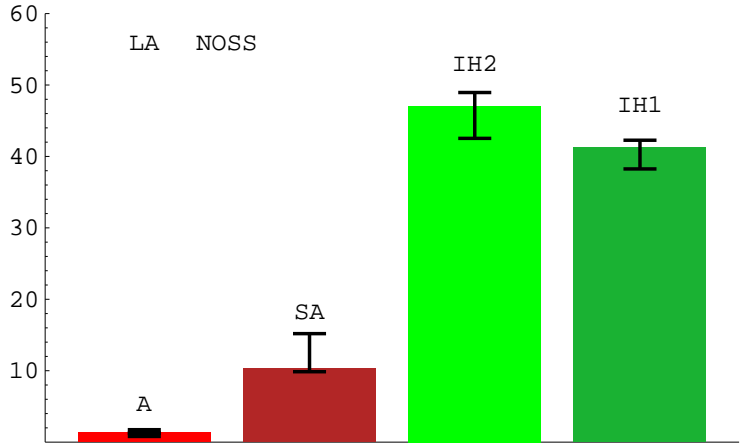,width=0.65\textwidth}}
\caption{Relative success rates for the LA solution, without see-saw. 
The sum of the rates has been normalized to 100. The results correspond to the 
default choice ${\cal I}=[0.5,2]$, and to the following values 
of $\lambda=\lambda'$: $0.2$, $0.2$, $0.25$, $0.3$
for the models ${\rm A_{NOSS}}$, ${\rm SA_{NOSS}}$, ${\rm IH_{(NOSS,II)}}$, 
and ${\rm IH_{(NOSS,I)}}$, respectively (in our notation there are no ${\rm H_{(NOSS,I)}}$, 
${\rm H_{(NOSS,II)}}$ models). 
The error bars represent the linear sum of the systematic error due to the
choice of ${\cal I}$ and the statistical error (see text).}
\label{flanoss}
\end{figure}

\begin{figure}[!h]
\centerline{\psfig{file=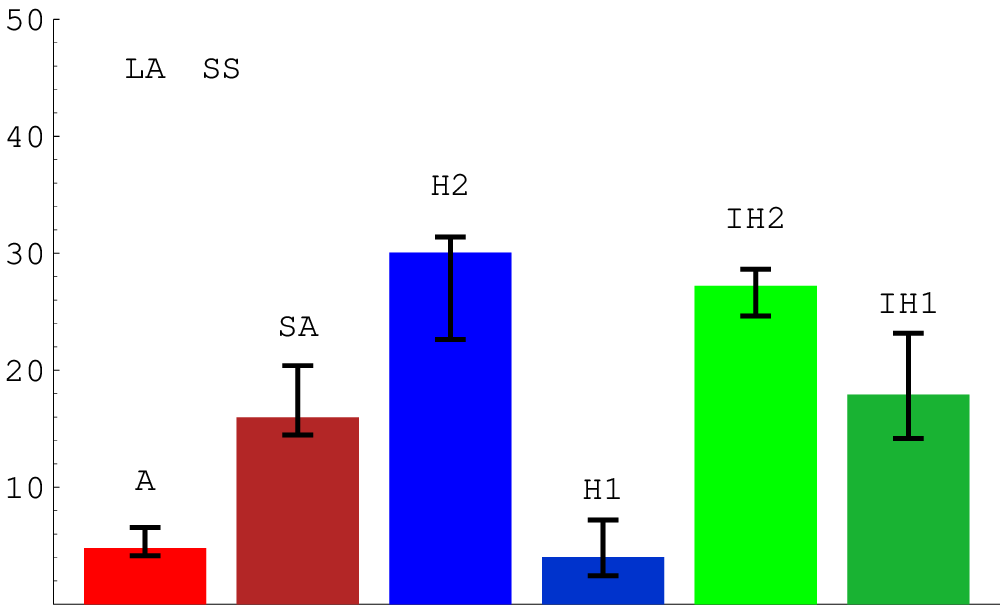,width=0.85\textwidth}}
\caption{Relative success rates for the LA solution, with see-saw. 
The sum of the rates has been normalized to 100. The results correspond to the 
default choice ${\cal I}=[0.5,2]$, and to the following values 
of $\lambda=\lambda'$: $0.2$, $0.3$, $0.35$, $0.5$, $0.15$, $0.2$ 
for the models ${\rm A_{SS}}$, ${\rm SA_{SS}}$, ${\rm H_{(SS,II)}}$, 
${\rm H_{(SS,I)}}$, ${\rm IH_{(SS,II)}}$ and ${\rm IH_{(SS,I)}}$, 
respectively. The error bars represent
the linear sum of the systematic error due to the
choice of ${\cal I}$ and the statistical error (see text).}
\label{flass}
\end{figure}

\begin{figure}[!t]
\vskip 1cm
\centerline{\psfig{file=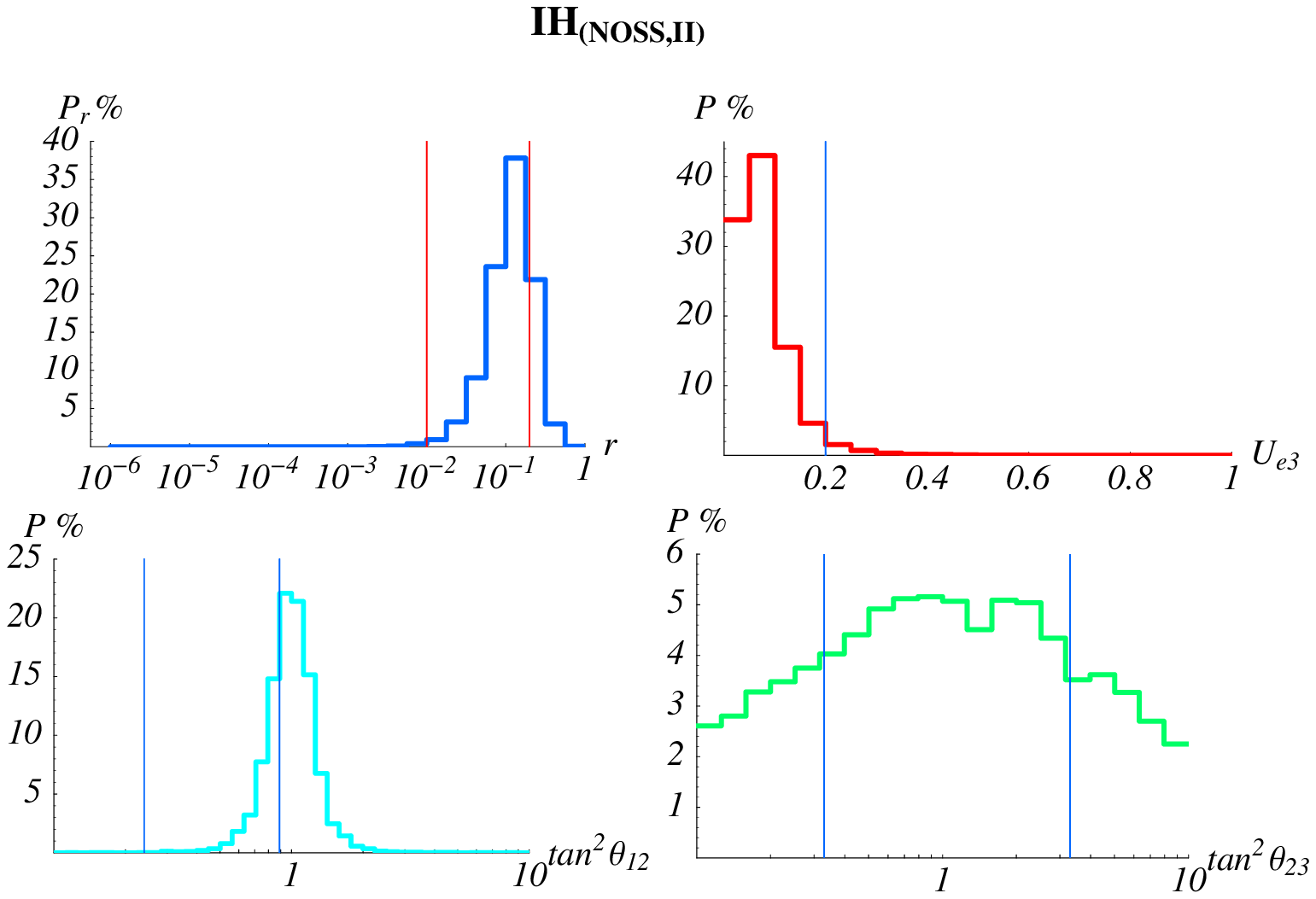,width=1\textwidth}}
\caption{Distributions for ${\rm IH_{(NOSS,II)}}$, ${\cal I}=[0.5,2]$,
$\lambda=\lambda'=0.25$, obtained with
10000 points ${\cal P}$.}
\label{fihnoss2}
\vspace*{1 cm}
\end{figure}

The success rates for the LA solutions are displayed in fig. \ref{flanoss}  and \ref{flass}, separately for the NOSS and SS
cases. The two sets of models have been individually normalized to give a total rate 100. Before normalization the total
success rates for NOSS and for SS were in the ratio 1.7:1 (see also appendix A).   Although the gaps between the rates of
different models are reduced compared to the LOW case, nevertheless a clear pattern emerges from these figures. The present
data are most easily described by the IH schemes in their NOSS version. Their performances are better by a factor of 10-30
with respect to the last classified, the anarchical models. The ability of the IH schemes in describing the data can be
appreciated from the distributions of the four observables, which, for 
${\rm IH_{(NOSS,II)}}$ and $\lambda=\lambda'=0.25$, are displayed  in fig. \ref{fihnoss2}.

\begin{figure}[!ht]
\vspace*{1.5cm}
\centerline{\psfig{file=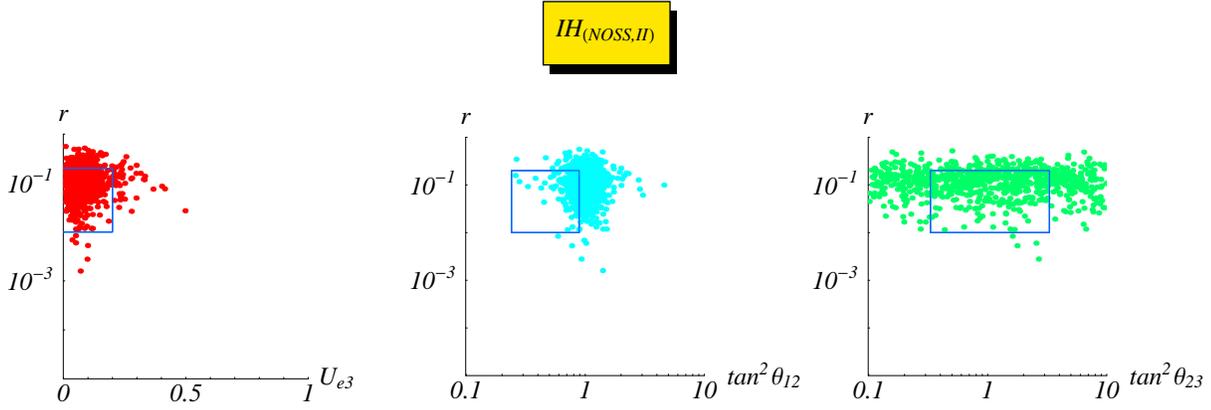,width=1.1\textwidth}}
\caption{Scatter plots in the planes $(U_{e3},r)$,
$(\tan^2\theta_{12},r)$ and $(\tan^2\theta_{23},r)$ for 
${\rm IH_{(NOSS,II)}}$,  
${\cal I}=[0.5,2]$, $\lambda=\lambda'=0.25$, obtained with
1000 points ${\cal P}$. The box shows the experimental window for LA.}
\label{nuvole1}
\end{figure}

The observables $r$ and $U_{e3}$ are strongly correlated. Actually, as discussed in ref. \cite{af02,isa1} and shown in table 2, in inversely
hierarchical models $U_{e3}$ is typically of order $r$. Therefore, once $\lambda$ and $\lambda'$ have been tuned to fit $r$,
this choice automatically provides a good fit  to $U_{e3}$. Moreover, similarly to the case of the LOW solution, $\tan^2
\theta_{12}$ is peaked around 1.  At present $\tan^2 \theta_{12}=1$ is excluded for the LA solution, but, thanks to the
width of the distribution,  the experimentally allowed window is sufficiently populated. The width of the distribution is
almost entirely dominated by the effect coming from the diagonalization of the charged lepton sector. Indeed, by turning off
the small parameters
$\lambda$ and $\lambda'$ in the mass matrix for the charged  leptons, we get a vanishing success rate for ${\rm
IH_{(NOSS,I)}}$, whereas the rate for ${\rm IH_{(NOSS,II)}}$ decreases by more than one  order of magnitude. It is worth
stressing that even a moderate further departure of the  window away from $\tan^2 \theta_{12}=1$ could drastically reduce
the success rates of the IH schemes.  Finally, the $\tan^2\theta_{23}$ distribution is rather flat, with a moderate peak in
the currently favoured interval. All the IH models, with or without see-saw, have distributions similar to those shown in
fig. \ref{fihnoss2}. In particular the $\tan^2\theta_{23}$ distribution of fig. \ref{fihnoss2}  is qualitatively common to
all U(1) models. This  reflects the fact that the large  angle $\theta_{23}$ is induced by the equal charges 
${\rm F}({\bar 5}_2)={\rm F}({\bar 5}_3)$, a feature shared by  all the models we have considered. To appreciate the
relevant correlations,  the distributions for the model ${\rm IH_{(NOSS,II)}}$ are also displayed  in fig. \ref{nuvole1}, as
scatter plots in the planes
$(U_{e3},r)$, $(\tan^2\theta_{12},r)$ and $(\tan^2\theta_{23},r)$.

\vskip 2cm

\begin{figure}[!t]
\vskip .5cm
\centerline{\psfig{file=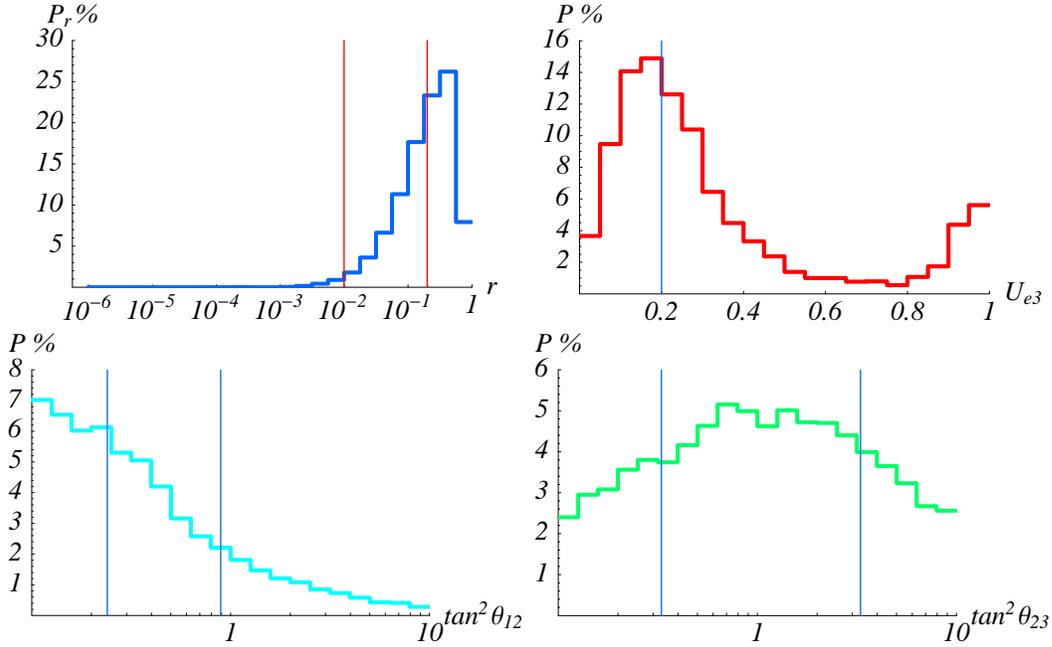,width=1\textwidth}}
\caption{Distributions for the semi-anarchical no-see-saw ${\rm SA_{NOSS}}$, ${\cal I}=[0.5,2]$,
$\lambda=\lambda'=0.2$, obtained with 10000 points ${\cal P}$.}
\label{fsanoss}
\end{figure}

Without see-saw mechanism, the next successful model is the semi-anarchical model SA, whose  distributions are displayed in fig. \ref{fsanoss} (in our
notation there are no ${\rm H_{NOSS,I(II)}}$ models). Compared to the IH case, the $\tan^2\theta_{12}$ distribution has no
pronounced peak. Possible shifts in the central value of
$\tan^2\theta_{12}$ would not drastically modify the results for the SA model. The $U_{e3}$ distribution is peaked around
$\lambda=0.2$ with tails that exceed the present experimental bound.  The $r$ distribution is centered near $r=1$. Finally,
the anarchical scheme  in its NOSS version is particularly disfavoured,  due to its tendency to predict $r$ close to 1 and
also due  to $U_{e3}$, that presents a broad distribution with a preferred value of about 0.5.

\begin{figure}[!h]
\centerline{\psfig{file=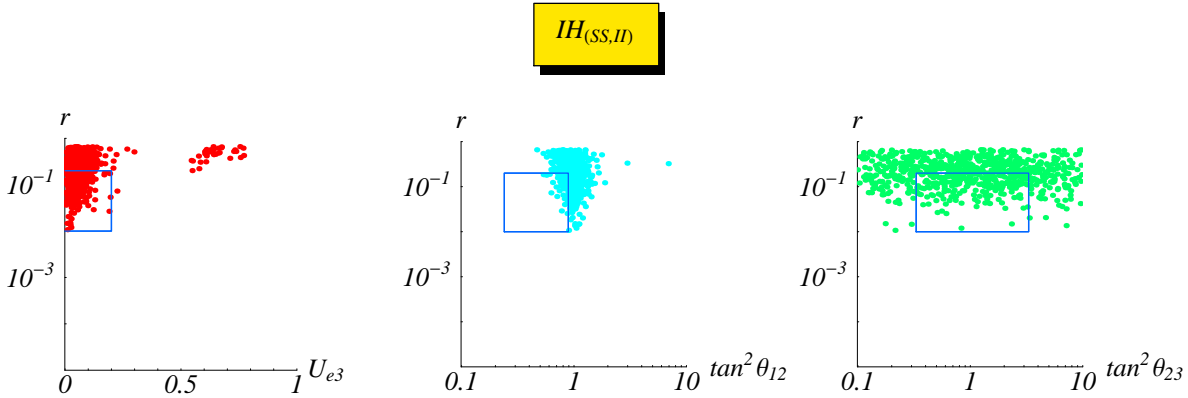,width=1.1\textwidth}}
\caption{Scatter plots in the planes $(U_{e3},r)$,
$(\tan^2\theta_{12},r)$ and $(\tan^2\theta_{23},r)$ for ${\rm IH_{(SS,II)}}$,  
${\cal I}=[0.5,2]$, $\lambda=\lambda'=0.15$, obtained with
1000 points ${\cal P}$. The box shows the experimental window for LA.}
\label{nuvole2}
\end{figure}

The overall picture changes significantly if the LA solution is realized in the context of the see-saw mechanism, as
illustrated in fig.
\ref{flass}.  The IH models are still rather successful. Compared to the NOSS case, the ${\rm IH_{SS}}$ models slightly
prefer  higher values of $r$ and, due to a smaller $\lambda=\lambda'$,  they have very narrow $\tan^2 \theta_{12}$
distributions. This can be seen from the scatter plots of fig. \ref{nuvole2}. The other distributions are similar to those
of figs. \ref{fihnoss2} and 
\ref{nuvole1}. We observe that while most of the points in fig. \ref{nuvole2}
are centered around $U_{e3}\approx O(\lambda^2)$, there is also a small
region clustered at $U_{e3}\approx 0.6$.

\begin{figure}[!h]
\centerline{\psfig{file=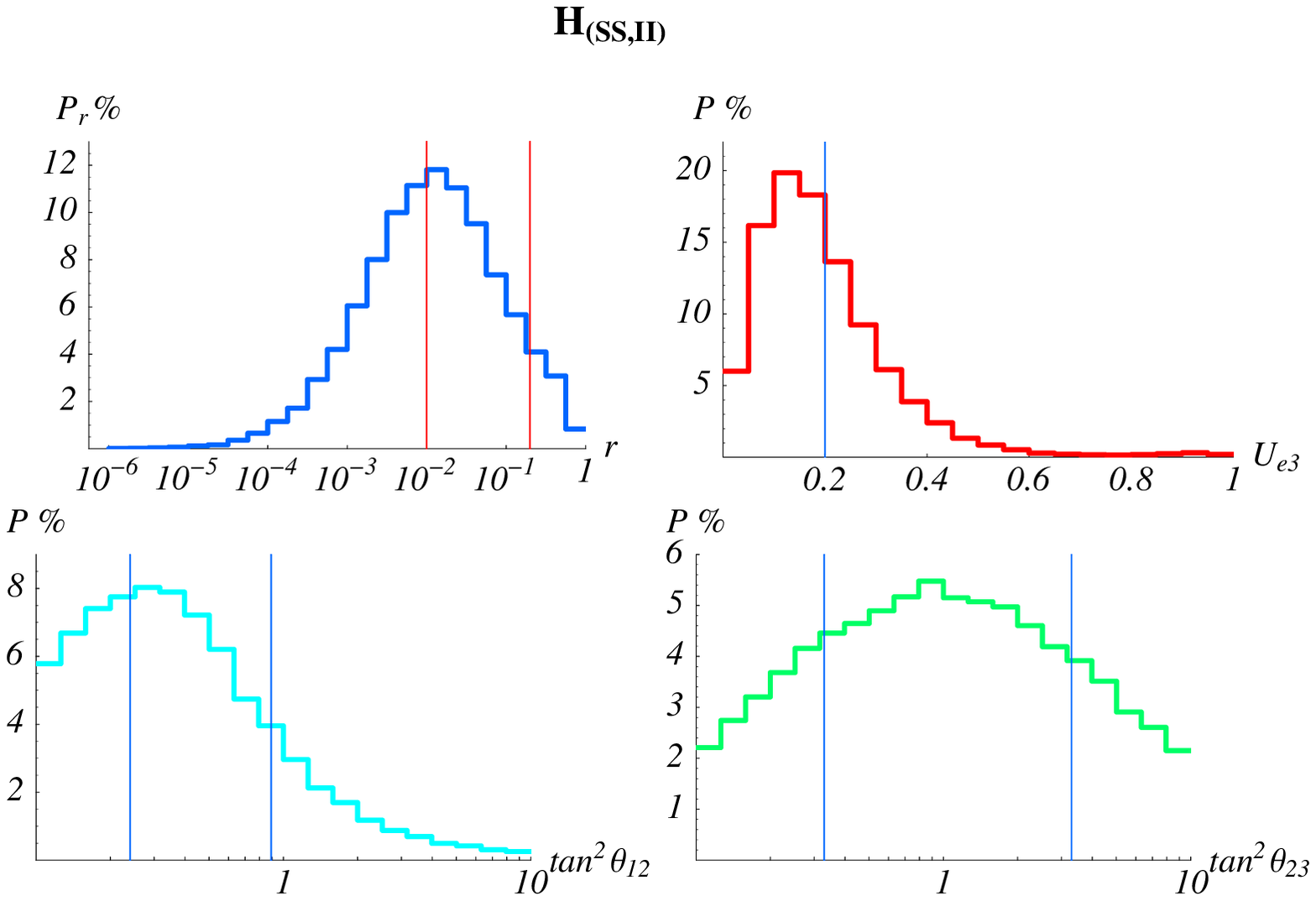,width=1\textwidth}}
\caption{Distributions for ${\rm H_{(SS,II)}}$, ${\cal I}=[0.5,2]$, 
$\lambda=\lambda'=0.35$, obtained with 50000 points ${\cal P}$.}
\label{fhss2}
\end{figure}

\begin{figure}[!h]
\centerline{\psfig{file=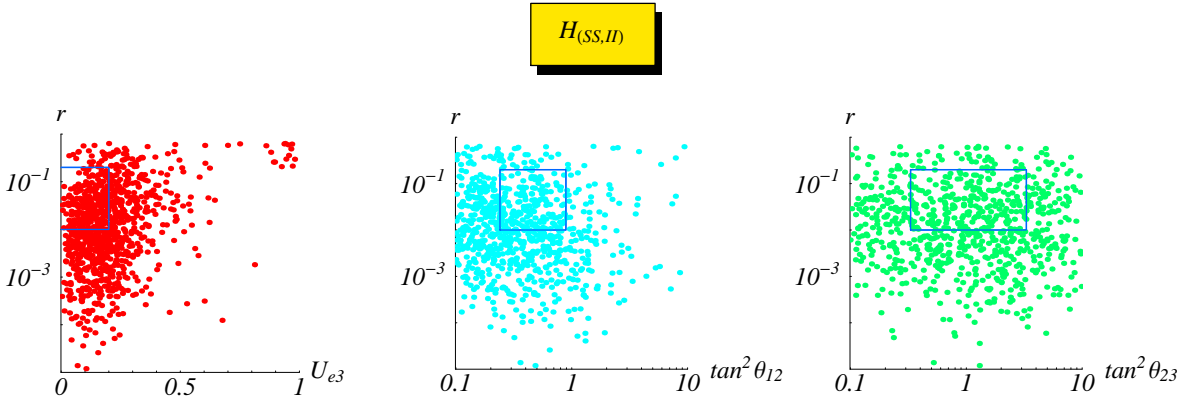,width=1.1\textwidth}}
\caption{Scatter plots in the planes $(U_{e3},r)$,
$(\tan^2\theta_{12},r)$ and $(\tan^2\theta_{23},r)$ for ${\rm H_{(SS,II)}}$,  
${\cal I}=[0.5,2]$, $\lambda=\lambda'=0.35$, obtained with
1000 points ${\cal P}$. The box shows the experimental window for LA.}
\label{nuvole3}
\end{figure}

Equally good or even better results are obtained by  the ${\rm H_{(SS,II)}}$ model, with distributions shown in fig. 
\ref{fhss2} and \ref{nuvole3}. We see that, at variance with the IH models, 
$\tan^2\theta_{12}$ is not spiky, which results in a  better stability of the model against variation of the experimental 
results. The preferred value of $r$ is close to the lower end of the experimental window. The $U_{e3}$ distribution is
nicely peaked around $\lambda^2$.

\begin{figure}[!h]
\centerline{\psfig{file=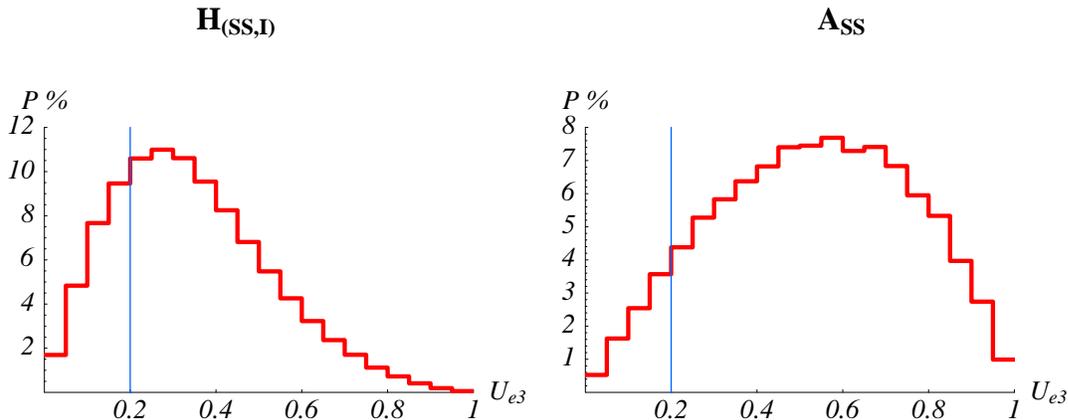,width=1\textwidth}}
\caption{$U_{e3}$ Distributions for ${\rm H_{(SS,I)}}$ ($\lambda=\lambda'=0.5$)
and ${\rm A_{SS}}$ ($\lambda=\lambda'=0.2$), ${\cal I}=[0.5,2]$, 
obtained with 50000 points ${\cal P}$.}
\label{ue3}
\end{figure}

The ${\rm SA_{SS}}$ model is significantly outdistanced from
${\rm H_{(SS,II)}}$, ${\rm IH_{(SS,II)}}$ and ${\rm IH_{(SS,I)}}$. It is particularly penalized by the $U_{e3}$
distribution, centered around $\lambda=0.3$. Finally, the least favoured models are ${\rm H_{(SS,I)}}$ and 
${\rm A_{SS}}$. The model ${\rm H_{(SS,I)}}$ fails both in $U_{e3}$ (see fig. \ref{ue3}), which tends to be  too large for
the preferred value of $\lambda=\lambda'=0.5$ and in 
$\tan^2\theta_{12}$. The ${\rm A_{SS}}$ model, as its NOSS version, suffers especially from the $U_{e3}$ distribution (see
fig. \ref{ue3}) which is roughly  centered at 0.5, with only few percent of the attempts falling within the present
experimental bound. A large $U_{e3}$ can be regarded as a specific prediction of anarchy and any possible improvement of the
bound on
$\vert U_{e3} \vert$ will wear away the already limited  success rate of the model.  The distributions in
$\tan^2\theta_{12}$ and $\tan^2\theta_{23}$ are equally broad and peaked around 1. Compared to the NOSS case, 
${\rm A_{SS}}$ has a better $r$ distribution, well located inside the  allowed window.

\vskip .5cm 
\vskip 1cm

As another criterion for evaluating the quality of a model, we address the issue of the stability
of the observables $O_i$ with respect to small fluctuations of the set of coefficients $\cal P$.
Notice that the random coefficients to be put in front of the powers of $\lambda, \lambda'$ stand for the
combined result of a fundamental theory of flavour, present at a certain scale $\Lambda$, 
and of an evolution from $\Lambda$ down to $m_Z$. 
It would thus be natural if the physical observables $O_i$ were stable under small perturbations
of the coefficients, $\Delta {\cal P}$, around a given successful representative set ${\cal P}_0$.   

\begin{figure}[!t]
\centerline{\psfig{file=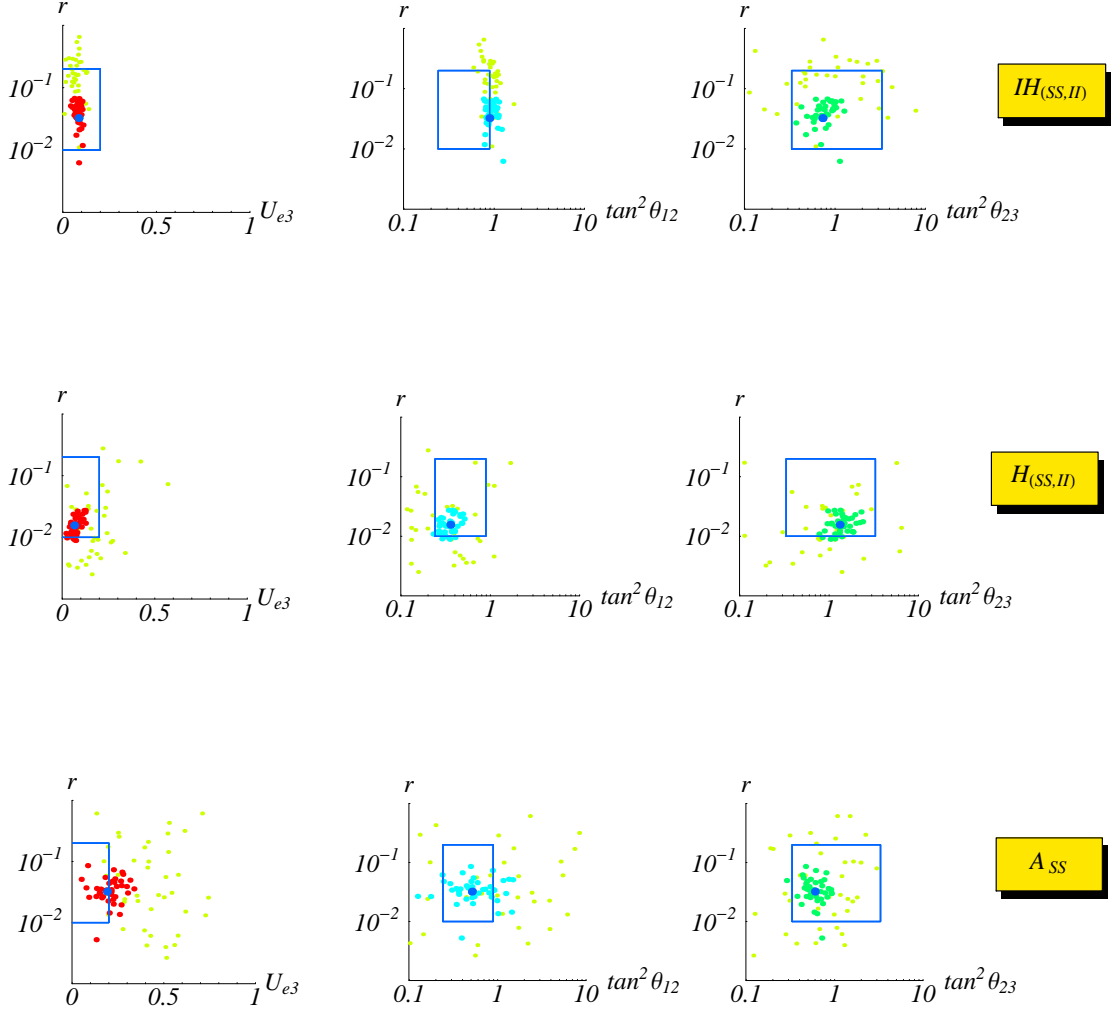,width=1.\textwidth}}
\caption{Results of the stability test in the planes $(U_{e3},r)$,
$(\tan^2\theta_{12},r)$ and $(\tan^2\theta_{23},r)$ for 
${\rm IH_{(SS,II)}}$ ($\lambda=\lambda'=0.15$), ${\rm H_{(SS,II)}}$ ($\lambda=\lambda'=0.35$)
and ${\rm A_{SS}}$ ($\lambda=\lambda'=0.2$), ${\cal I}=[0.5,2]$ (see text).}
\label{stabss}
\end{figure}
\vskip 0.3cm

\begin{figure}[!h]
\centerline{\psfig{file=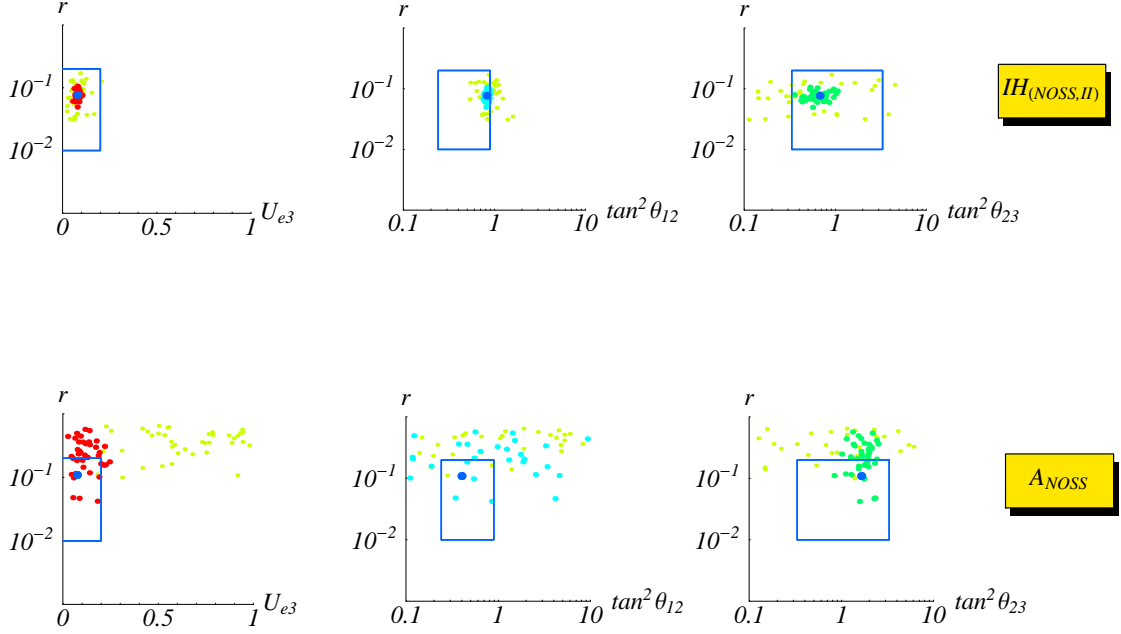,width=1.\textwidth}}
\caption{Results of the stability test in the planes $(U_{e3},r)$,
$(\tan^2\theta_{12},r)$ and $(\tan^2\theta_{23},r)$ for ${\rm IH_{(NOSS,II)}}$ ($\lambda=\lambda'=0.25$)
and ${\rm A_{NOSS}}$ ($\lambda=\lambda'=0.2$),
${\cal I}=[0.5,2]$ (see text).}
\label{stabnoss}
\end{figure}

This is illustrated for the seesaw case in fig. \ref{stabss}, where we compare the models 
${\rm IH_{(SS,II)}}$, ${\rm H_{(SS,II)}}$ and ${\rm A_{SS}}$.
The blue dot refers to the observables $O_i$ which follow from a typical successful configuration for LA,
${\cal P}_0$. 
The 40 points in red, light blue and green are the result of adding to ${\cal P}_0$
random perturbations $\Delta {\cal P}$ with $|\Delta {\cal P}|=|{\cal P}|/10$ and random phases.
The yellow dots correspond to $|\Delta {\cal P}|=|{\cal P}|/2$.
As appears from the scatter plots, $\rm IH_{(SS,II)}$ is very stable: as already argued in the above discussion,
actually it is even dangerously stable with respect to the prediction for $\tan^2\theta_{12}$.
If models with hierarchy display a sufficient degree of stability,
anarchical ones are much less stable, in particular with respect to the predictions for $U_{e3}$ and  
$\tan^2 \theta_{12}$. 
As shown in fig. \ref{stabnoss}, the non see-saw case enhances these features: $\rm IH_{(NOSS,II)}$
is extremely stable, while $\rm A_{NOSS}$ is higly unstable. For the latter, in particular, 
even with the small fluctuation considered in the figures, $\tan^2\theta_{12}$ spans from .1 to 10.   
Thus, the criterion of stability supports the same ratings of models previously obtained by only considering
total rates and distributions.

\vskip .5cm 

As a general comment we observe that our results are rather stable with  respect to the choice of the interval ${\cal I}$.
With only one exception, namely the crossing between $H_{(SS,I)}$ and $A_{SS}$, the relative position of the different
models, according to their ability in describing the data, does not vary when we shrink ${\cal I}$, or when we extend it to
cover the full circle of radius 1
\footnote{The results for ${\cal I}=[0,n]$ are independent on $n$, since changing $n$ amounts to perform a renormalization
of all mass matrices by a common overall scale, which is not felt by the observables we have used in our analysis.}. This
stability would be partially upset by restricting to the case of real coefficients ${\cal P}$. In that case the relative
rates are comparable to those obtained in the complex case only for sufficiently wide intervals ${\cal I}$, typically 
${\cal I}\approx [-1/\sqrt{\lambda},-\sqrt{\lambda}] \cup  [\sqrt{\lambda},1/\sqrt{\lambda}]$. If we further squeeze ${\cal I}$
around $\pm 1$ the rates of all models tend to zero.
In the real case, the distribution of $\tan^2 \theta_{23}$ is very sensitive
to the width of the extraction interval $\cal I$. Indeed, in the flavour 
symmetry basis, charged leptons and light neutrinos mass matrices are both 
diagonalised by a 23 mixing angle which tends to $\pi/4$ when $\cal I$
is squeezed to $1$. As an effect, in this limit the distribution of 
$\theta_{23}$ is almost empty around $\pi/4$ and presents two peaks at 
$0$ and $\pi/2$. On the contrary, in the complex case the distribution of 
$\theta_{23}$ is quite insensitive to the width of $\cal I$ thanks to the 
smearing effect of the phases.
%
%

\vspace{0.5cm}
\section{Conclusion}

If a large gap between the solar and atmospheric frequencies for neutrino oscillations were finally to be established by
experiment then this fact would immediately suggest that neutrino masses are hierarchical,
similar to quark and charged lepton masses. In fact very small mass squared differences among nearly degenerate neutrino
states are difficult to obtain and make stable under running in the absence of an ad hoc symmetry. Moreover, the
presence of very small mixing angles would also indicate a hierarchical pattern. At present, the SA solution of solar
neutrino oscillations is disfavoured so that 2 out of 3 mixing angles appear to be large and only one appears to be small,
although the present limit is not terribly constraining. However, in the case of the LA solution for solar neutrino
oscillations, the ratio $r$ of the solar to atmospheric frequencies is not so small and it has been suggested that possibly
this solution does not require a symmetry to be generated. In the anarchical framework the smallness of
$U_{e3}$, assumed not too pronounced, and that of $r$, are considered as accidental (the see-saw mechanism helps
in this respect because the product of 3 matrices sizeably broadens the $r$ distribution). In the present study we have
examined in quantitative terms the relative merits of anarchy and of different implementations of hierarchy in
reproducing the observed features of the LA solution. For our analysis we have adopted the framework of 
SU(5)$\times$U(1)$_{\rm F}$ which is flexible enough, by suitable choices of the flavour charges, to reproduce all interesting types of hierarchy
and also of anarchy. This framework allows a statistical comparison of the different schemes under, as far as possible, homogeneous
conditions. 
The rating of models in terms of
their statistical rates of success is clearly a questionable procedure.
After all Nature does not proceed at random and a particular mass pattern
that looks odd could arise due to some deep dynamical reason. However,
the basis for anarchy as a possible description of the LA solution can
only be formulated in statistical terms. Therefore it is interesting to
compare anarchy versus hierarchy on the same grounds.
We have considered models both with normal and with inverse hierarchies, with and without see-saw, with one or
two flavons, and compared them with the case of anarchy and of semi-anarchy (models where there is no structure in 2-3 but
only in 2-3 vs 1). The stability of our results has been tested by considering different options for the statistical
procedure and also by studying the effect on each type of model of small parameter changes. 

Our conclusion is that, for the LA solution, the range of $r$ and the small upper
limit on
$U_{e3}$  are sufficiently constraining to make anarchy neatly disfavoured with respect to models
with built-in hierarchy. If only neutrinos are considered, one might counterargue that hierarchical models have
at least one more parameter than anarchy, in our case the parameter $\lambda$. However, if one looks at quarks and
leptons together, as in the GUT models that we consider, then the same parameter that plays the role of an order parameter
for the CKM matrix, for example, the Cabibbo angle, can be successfully used to reproduce also the LA hierarchy. On the one
hand, it is interesting that the amount of hierarchy needed for the LA solution is just a small power of the Cabibbo angle.
On the other hand, if all fermion masses and mixings are to be reproduced, a similar parameter is also needed in anarchical
models, where all the quark and lepton mass structure
arises from the charges of $\Psi^{10}_i$. In comparison to anarchy even the limited amount of
structure present in semi-anarchical models already improves the performance a lot. And the advantage is further
increased when more structure is added as in inverse hierarchy models, or models with normal hierarchy and automatic
suppression of the 23 determinant. In the see-saw case all these types of hierarchical models have comparable rates of
success (except for
${\rm H_{(SS,I)}}$). In the non-see-saw versions of inverse hierarchy the performance is even better. An experimental
criterion that could eventually decide between normal and inverse hierarchy models is the closeness of the solar angle
$\theta_{12}$ to its maximal value. If the data moves away from $\pi/4$ the probability of inverse hierarchy will rapidly
drop in comparison to hierarchical models. 
\vspace{0.5cm}

\section*{Acknowledgments} We would like to thank Francesco Vissani for useful discussions and for 
partecipating in the early stages of the present work. 
F.F. and I.M. thank the CERN TH divison for its hospitality
during july and august 2002, when this project was developed. This work is partially
supported by the European Programs HPRN-CT-2000-00148 and HPRN-CT-2000-00149.

\newpage

\section*{Appendix: raw data}
In this section we list the results of our numerical analysis, for the case of complex random
coefficients ${\cal P}$. In particular we detail the dependence of the success rates on
the size of the window ${\cal I}$ that specifies the absolute value of the coefficients ${\cal P}$.
In most cases the `systematic' error due to the choice of ${\cal I}$ is larger than the statistical 
error. The latter is given by $\delta P/P=1/\sqrt{N_s}$, where $P=100\cdot N_s/N$ is the success rate and 
$N_s$ is the number of successes in $N$ trials. We chose $N=10000$, $30000$, $50000$ respectively,
for the LOW solution, for LA without see-saw and for LA with see-saw. When fitting the LOW solution,
the results for the anarchical and semi-anarchical models do not depend on $\lambda$ or $\lambda'$.
\begin{table}[!h]
\caption{Success rates for the LOW solution, with the see-saw mechanism.
\label{tab3}}
\begin{center}
\begin{tabular}{|c|c|c|c|c|c|}
\hline
\multicolumn{2}{|c|}{}&\multicolumn{4}{|c|}{}\\
\multicolumn{2}{|c|}{}&\multicolumn{4}{|c|}{\tt success rate}\\
\multicolumn{2}{|c|}{}&\multicolumn{4}{|c|}{}\\
\hline
& & & & &\\
{\tt model}& $\lambda(=\lambda')$& ${\cal I}=[0.5,2]$ & ${\cal I}=[0.8,1.2]$ & ${\cal I}=[0.95,1.05]$ & ${\cal I}=[0,1]$\\
& & & & &\\
\hline
& & & & &\\
$A_{SS}$& - & $<0.01$ & $<0.01$& $<0.01$& $<0.01$\\
& & & & &\\
\hline
& & & & &\\
$SA_{SS}$& - & $<0.01$ & $<0.01$& $<0.01$& $<0.01$\\
& & & & &\\
\hline
& & & & &\\
$H_{(SS,II)}$& 0.1 & $1.7\pm 0.1$ & $2.8\pm 0.2$ & $2.8\pm 0.2$ & $1.6\pm 0.1$\\
& & & & &\\
\hline
& & & & &\\
$H_{(SS,I)}$& 0.15 & $14.4\pm 0.4$ & $21.6\pm 0.5$ & $23.0\pm 0.5$ & $7.6\pm 0.3$\\
& & & & &\\
\hline
& & & & &\\
$IH_{(SS,II)}$& 0.03& $41.6\pm 0.6$ & $30.3\pm 0.6$ & $29.0\pm 0.5$ & $31.2\pm 0.6$ \\
& & & & &\\
\hline
& & & & &\\
$IH_{(SS,I)}$& 0.04& $42.9\pm 0.7$ & $32.8\pm 0.6$ & $31.6\pm 0.6$ & $32.0\pm 0.6$\\
& & & & &\\
\hline
\end{tabular}
\end{center}
\end{table}
\\[0.1cm]
\begin{table}[!h]
\caption{Success rates for the LOW solution, without the see-saw mechanism.
\label{tab3noss}}
\begin{center}
\begin{tabular}{|c|c|c|c|c|c|}
\hline
\multicolumn{2}{|c|}{}&\multicolumn{4}{|c|}{}\\
\multicolumn{2}{|c|}{}&\multicolumn{4}{|c|}{\tt success rate}\\
\multicolumn{2}{|c|}{}&\multicolumn{4}{|c|}{}\\
\hline
& & & & &\\
$A_{NOSS}$&  - & $<0.01$ & $<0.01$& $<0.01$& $<0.01$\\
& & & & &\\
\hline
& & & & &\\
$SA_{NOSS}$& - & $<0.01$ & $<0.01$& $<0.01$& $<0.01$\\
& & & & &\\
\hline
& & & & &\\
$IH_{(NOSS,II)}$& 0.05 & $43.4\pm 0.7$ & $33.3\pm 0.6$ & $32.1\pm 0.6$ & $42.9\pm 0.7$\\
& & & & &\\
\hline
& & & & &\\
$IH_{(NOSS,I)}$& 0.06 & $46.1\pm 0.7$ & $36.5\pm 0.6$ & $35.5\pm 0.6$ & $38.8\pm 0.6$\\
& & & & &\\
\hline
\end{tabular}
\end{center}
\end{table}
\begin{table}[!h]
\caption{Success rates for the LA solution, without see-saw mechanism. 
\label{tab4noss}}
\begin{center}
\begin{tabular}{|c|c|c|c|c|c|}
\hline
\multicolumn{2}{|c|}{}&\multicolumn{4}{|c|}{}\\
\multicolumn{2}{|c|}{}&\multicolumn{4}{|c|}{\tt success rate}\\
\multicolumn{2}{|c|}{}&\multicolumn{4}{|c|}{}\\
\hline
& & & & &\\
$A_{NOSS}$&  0.2 & $0.33\pm 0.03$ & $0.23\pm 0.03$& $0.25\pm 0.03$& $0.26\pm 0.03$\\
& & & & &\\
\hline
& & & & &\\
$SA_{NOSS}$& 0.2 & $2.5\pm 0.1 $ & $2.9\pm 0.1$& $2.6\pm 0.1$& $2.5\pm 0.1$\\
& & & & &\\
\hline
& & & & &\\
$IH_{(NOSS,II)}$& 0.25 & $11.4\pm 0.2$ & $8.8\pm 0.2$ & $8.2\pm 0.2$ & $10.2\pm 0.2$\\
& & & & &\\
\hline
& & & & &\\
$IH_{(NOSS,I)}$& 0.3 & $10.0\pm 0.2$ & $7.9\pm 0.2$ & $7.7\pm 0.2$ & $8.4\pm 0.2$\\
& & & & &\\
\hline
\end{tabular}
\end{center}
\end{table}
\\[0.1cm]
\begin{table}[!h]
\caption{Success rates for the LA solution, with see-saw mechanism. 
\label{tab4}}
\begin{center}
\begin{tabular}{|c|c|c|c|c|c|}
\hline
\multicolumn{2}{|c|}{}&\multicolumn{4}{|c|}{}\\
\multicolumn{2}{|c|}{}&\multicolumn{4}{|c|}{\tt success rate}\\
\multicolumn{2}{|c|}{}&\multicolumn{4}{|c|}{}\\
\hline
& & & & &\\
{\tt model}& $\lambda(=\lambda')$& ${\cal I}=[0.5,2]$ & ${\cal I}=[0.8,1.2]$ & ${\cal I}=[0.95,1.05]$ & ${\cal I}=[0,1]$\\
& & & & &\\
\hline
& & & & &\\
$A_{SS}$& 0.2 & $0.69\pm 0.04$ & $0.62\pm 0.04$& $0.63\pm 0.04$& $0.63\pm 0.04$\\
& & & & &\\
\hline
& & & & &\\
$SA_{SS}$& 0.3 & $2.30\pm 0.07$ & $2.12\pm 0.07$& $2.07\pm 0.06$& $1.99\pm 0.06$\\
& & & & &\\
\hline
& & & & &\\
$H_{(SS,II)}$& 0.35 & $4.33\pm 0.09$ & $4.35\pm 0.09$ & $4.34\pm 0.09$ & $2.36\pm 0.07$\\
& & & & &\\
\hline
& & & & &\\
$H_{(SS,I)}$& 0.5 & $0.58\pm 0.03$ & $0.88\pm 0.04$ & $0.97\pm 0.04$ & $0.27\pm 0.02$\\
& & & & &\\
\hline
& & & & &\\
$IH_{(SS,II)}$& 0.15& $3.92\pm 0.09$ & $3.97\pm 0.09$ & $4.06\pm 0.09$ & $2.57\pm 0.07$ \\
& & & & &\\
\hline
& & & & &\\
$IH_{(SS,I)}$& 0.2& $2.58\pm 0.07$ & $2.24\pm 0.07$ & $2.07\pm 0.06$ & $2.26\pm 0.07$\\
& & & & &\\
\hline
\end{tabular}
\end{center}
\end{table}

\vfill
\newpage
\section{Addendum}

Since the completion of our analysis new important experimental results have been published, in particular the first data
from KamLAND \cite{Kam} and the new results from SNO \cite{SNOnew}. The main new information, of special relevance for the
present analysis, is that the LOW solution is now discarded and the window corresponding to the LA solution has considerably
narrowed. In particular the value of $r$ has settled around $r\approx 1/35$ and the solar angle $\theta_{12}$ is now more than
$5\sigma$ away from maximal. For example, in ref. \cite{malt} from a comprehensive analysis
of all the available data the following $3\sigma$ window was obtained :
\be
\begin{array}{l} 0.018 < r < 0.053\\
\vert U_{e3} \vert < 0.23\\ 0.30 <\tan^2\theta_{12}<0.64\\ 0.45 <\tan^2\theta_{23}<2.57
\end{array}
\label{lanew}
\ee
These values are to be compared with those in eq. (\ref{la}), although in eq. (\ref{lanew}) we have a $3\sigma$ window
according to ref. \cite{malt} while the indicative ranges in eq. (\ref{la}) were less precisely defined. 
These new experimental
developments have a large enough impact on the present approach to make it certainly worthwhile and timely to update the
results of our analysis taking the new information into account. As discussed in our paper, the deviation of the solar
mixing angle from the maximal value is expected to produce a pronounced decrease of the success rate of the IH models. 
Also
the considerable narrowing of the allowed $r$ range (the upper part of the LA solution is now disfavoured) should have
important consequences on the relative effectiveness of A, SA, H and IH models. It is interesting to precisely quantify
these issues and this addendum is devoted to this task.

Given the new values in eq. (\ref{lanew}), the values of $\lambda$ and $\lambda'$ have been reoptimized and in some cases
their values are somewhat changed with respect to the previous numbers, as shown in Tables 7 and 8, which replace Tables 5
and 6.
The values of the charges for the representation 10 have also been modified for those cases where a rather large value of
$\lambda$ follows from the optimization (in order to maintain the correct amount of quark and lepton mass hierarchies), as
shown in Table 9 that replaces Table 1. Since the neutrino mixing parameters are completely independent on
the 10 charges, this change is only important for a
better fit to quark and charged lepton masses and mixings.  

The new versions of figs. 2 and 3 that describe the success rates separately for the NOSS and SS cases are presented in figs.
13 and 14. The main qualitative difference between the
new and the old rates is that indeed, both in the NOSS and the SS cases, a large drop in the IH relative success rate is
immediately apparent. This is due not only to the smaller upper limit on the solar mixing angle
$\theta_{12}$ but also to the smaller value of $r$ (in fact, we recall from Table 2 that $r$ and
$\theta_{13}$ are expected to be of the same order in IH models).  
From the updated
histograms in figs. 13 and 14 we see that normal hierarchy models (with two oppositely charged flavons ${\rm H_{II}}$) are neatly
preferred over anarchy and inverse hierarchy in the context of these SU(5)$\times$U(1) models. In particular, in the SS
case, the ${\rm H_{II}}$ models with normal hierarchy and suppressed 23 sub-determinant are clearly preferred. 
We recall that for the chosen charge values the ${\rm H_{II}}$ model is of the lopsided type.

In conclusion, with all the limitations of the present approach, it is interesting that the hierarchical models are
preferred over anarchy and inverse hierarchy. In particular in the case of see-saw dominance the preferred models are those
with natural suppression of the 23 subdeterminant (e.g those of the lopsided type or those with dominance of a light
right-handed eigenvalue) which indeed provide a simple natural solution of all known constraints.
\vspace{0.4cm}
\begin{table}[!h]
\caption{Success rates for the LA solution, without see-saw mechanism. \label{tab44noss}}
\begin{center}
\begin{tabular}{|c|c|c|c|c|c|}
\hline
\multicolumn{2}{|c|}{}&\multicolumn{4}{|c|}{}\\
\multicolumn{2}{|c|}{}&\multicolumn{4}{|c|}{\tt success rate}\\
\multicolumn{2}{|c|}{}&\multicolumn{4}{|c|}{}\\
\hline& & & & &\\
{\tt model}& $\lambda(=\lambda')$& ${\cal I}=[0.5,2]$ & ${\cal I}=[0.8,1.2]$ & 
${\cal I}=[0.95,1.05]$ & ${\cal I}=[0,1]$\\
& & & & &\\
\hline& & & & &\\
$A_{NOSS}$&  0.2 & $0.02\pm 0.01$ & $0.02\pm 0.01$& $0.02\pm 0.01$& $0.03\pm 0.01$\\
& & & & &\\
\hline& & & & &\\
$SA_{NOSS}$& 0.2 & $0.44\pm 0.03 $ & $0.74\pm 0.04$ & $0.68\pm 0.04$& $0.45\pm 0.03$\\
& & & & &\\
\hline& & & & &\\
$IH_{(NOSS,II)}$& 0.25 & $0.108\pm 0.015$ & $0.106\pm 0.015$ & $0.068\pm 0.012$ & $0.190\pm 0.019$\\
& & & & &\\
\hline& & & & &\\
$IH_{(NOSS,I)}$& 0.25 & $0.15\pm 0.02$ & $0$ & $0$ & $0.29\pm 0.02$\\
& & & & &\\
\hline
\end{tabular}
\end{center}
\end{table}
\vfill\newpage
\begin{table}[!h]
\caption{Success rates for the LA solution, with see-saw mechanism. \label{tab44}}
\begin{center}
\begin{tabular}{|c|c|c|c|c|c|}
\hline
\multicolumn{2}{|c|}{}&\multicolumn{4}{|c|}{}\\
\multicolumn{2}{|c|}{}&\multicolumn{4}{|c|}{\tt success rate}\\
\multicolumn{2}{|c|}{}&\multicolumn{4}{|c|}{}\\
\hline& & & & &\\
{\tt model}& $\lambda(=\lambda')$& ${\cal I}=[0.5,2]$ & ${\cal I}=[0.8,1.2]$ & 
${\cal I}=[0.95,1.05]$ & ${\cal I}=[0,1]$\\
& & & & &\\
\hline
& & & & &\\
$A_{SS}$& 0.2 & $0.15 \pm 0.02$ & $0.11\pm 0.02$& $0.12\pm 0.02$& $0.17\pm 0.02$\\
& & & & &\\
\hline
& & & & &\\
$SA_{SS}$& 0.25 & $0.53\pm 0.03$ & $0.38\pm 0.03$& $0.39\pm 0.03$& $0.46\pm 0.03$\\
& & & & &\\
\hline
& & & & &\\
$H_{(SS,II)}$& 0.35 & $1.041 \pm 0.023$ & $1.027\pm 0.032$ & $0.986\pm 0.031$ & $0.532\pm 0.023$\\
& & & & &\\
\hline
& & & & &\\
$H_{(SS,I)}$& 0.45 & $0.099\pm 0.010$ & $0.061\pm 0.008$ & $0.016\pm 0.004$ & $0.058\pm 0.008$\\
& & & & &\\
\hline
& & & & &\\
$IH_{(SS,II)}$& 0.45 & $0.033\pm 0.006$ & $0.007\pm 0.003$ & $0.011\pm 0.003$ & $0.092\pm 0.006$ \\
& & & & &\\
\hline
& & & & &\\
$IH_{(SS,I)}$& 0.25 & $0.020\pm 0.003$ & $0.013\pm 0.004$ & $0.006 \pm 0.003$ & $0.035\pm 0.004$\\
& & & & &\\
\hline
\end{tabular}
\end{center}
\end{table}
\vfill\newpage
\vspace{0.1cm}
\begin{table}[!h]
\caption{Models and their flavour charges. \label{updt_tab1}}
\vspace{0.4cm}
\begin{center}
\begin{tabular}{|c|c|c|c|c|}
\hline 
& & & & \\ 
{\tt Model}& ${{\Psi_{10}}}$ & ${\Psi_{\bar 5}}$ & ${{\Psi_1}}$ & ${(H_u,H_d)}$ \\ 
& & & & \\
\hline
\hline
 & & & & \\ 
{\tt Anarchical ($A$)}& (3,2,0)& (0,0,0) & (0,0,0) & (0,0)\\ 
& & & & \\
\hline
 & & & & \\ 
{\tt Semi-Anarchical ($SA$)}& (2,1,0) & (1,0,0) & (2,1,0) & (0,0) \\ 
& & & & \\
\hline
\hline 
& & & & \\
{\tt Hierarchical ($H_{I}$)}& (6,4,0)& (2,0,0) & (1,-1,0) & (0,0)\\ 
& & & & \\
\hline
& & & & \\
{\tt Hierarchical ($H_{II}$)}& (5,3,0)& (2,0,0) & (1,-1,0) & (0,0)\\ 
& & & & \\
\hline
& & & & \\ 
{\tt Inversely Hierarchical ($IH_{I}$)}& (3,2,0) & (1,-1,-1)& (-1,+1,0)& (0,+1) \\
& & & & \\
\hline& & & & \\ 
{\tt Inversely Hierarchical ($IH_{II}$)}& (6,4,0) & (1,-1,-1)& (-1,+1,0)& (0,+1) \\
& & & & \\
\hline
\end{tabular}
\end{center}
\end{table}
\vfill\newpage

\begin{figure}[!t]
\vskip .5 cm
\centerline{  \psfig{file=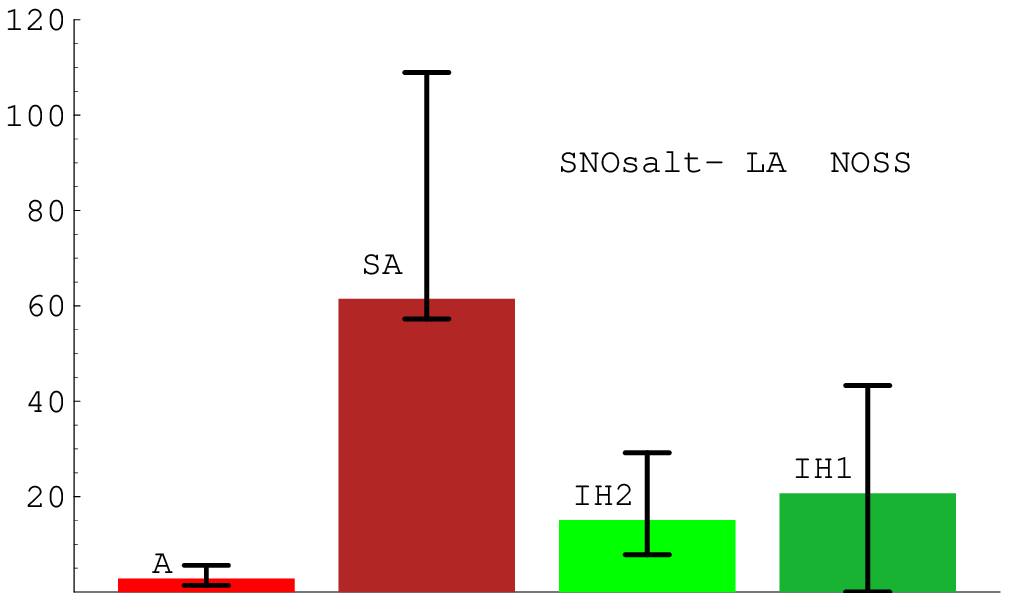,width=0.7 \textwidth}   }
\caption{Relative success rates for the LA solution, without see-saw. 
The sum of the rates has been normalized to 100. The results correspond to the 
default choice ${\cal I}=[0.5,2]$, and to the following values 
of $\lambda=\lambda'$: $0.2$, $0.2$, $0.25$, $0.25$
for the models ${\rm A_{NOSS}}$, ${\rm SA_{NOSS}}$, ${\rm IH_{(NOSS,II)}}$, 
and ${\rm IH_{(NOSS,I)}}$, respectively (in our notation there are no ${\rm H_{(NOSS,I)}}$, 
${\rm H_{(NOSS,II)}}$ models). 
The error bars represent the linear sum of the systematic error due to the
choice of ${\cal I}$ and the statistical error (see text).}
\label{updt_barlanoss}
\end{figure}
\begin{figure}[!h]
\centerline{   \psfig{file=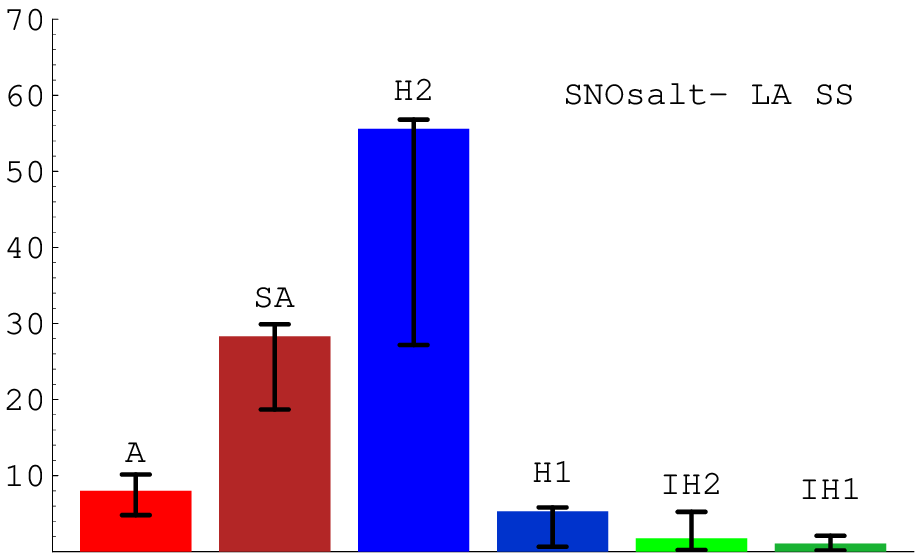,width=0.7 \textwidth}  }
\caption{Relative success rates for the LA solution, with see-saw. 
The sum of the rates has been normalized to 100. The results correspond to the 
default choice ${\cal I}=[0.5,2]$, and to the following values 
of $\lambda=\lambda'$: $0.2$, $0.25$, $0.35$, $0.45$, $0.45$, $0.25$ 
for the models ${\rm A_{SS}}$, ${\rm SA_{SS}}$, ${\rm H_{(SS,II)}}$, 
${\rm H_{(SS,I)}}$, ${\rm IH_{(SS,II)}}$ and ${\rm IH_{(SS,I)}}$, 
respectively. The error bars represent
the linear sum of the systematic error due to the
choice of ${\cal I}$ and the statistical error (see text).}
\label{updt_barlass}
\end{figure}

\end{document}